\documentstyle[preprint,aps,version2]{revtex}
\begin{document}
\draft
\begin{title}
Fermion-Spin Transformation to Implement the Charge-Spin
Separation
\end{title}

\author{Shiping Feng}
\begin{instit}
International Centre for Theoretical Physics, P. O. Box 586,
34100 Trieste, Italy, and \\
Department of Physics, Beijing Normal University, Beijing
100875, China \\
\end{instit}
\author{ Z.B. Su}
\begin{instit}
Institute of Theoretical Physics, Chinese Academy of Sciences,
Beijing 100080, China, and\\
International Centre for Theoretical Physics, P. O. Box 586,
34100 Trieste, Italy\\
\end{instit}
\author{L.Yu}
\begin{instit}
International Centre for Theoretical Physics, P. O. Box 586,
34100 Trieste, Italy, and \\
Institute of Theoretical Physics, Chinese Academy of Sciences,
Beijing 100080, China\\
\end{instit}


\begin{abstract}
A novel approach, the fermion-spin transformation to implement the
charge-spin separation, is developed to study the low-dimensional
$t$-$J$ model. In this approach, the charge and spin degrees of
freedom of the physical electron are separated, and the charge
degree of freedom is represented by a spinless fermion while the
spin degree of freedom is represented by a {\it hard-core boson}.
The on-site local constraint for single occupancy is satisfied
even in the mean-field approximation
and the sum rule for the physical electron is obeyed.
This approach can be applied to both one and
two-dimensional systems. In the one-dimensional case,
the spinon as well as the physical electron
behaves like Luttinger liquids. We have
obtained a gapless charge and spin excitation spectrum, a good
ground state energy, and a reasonable
electron-momentum distribution within the mean-field approximation.
The correct exponents of the correlation functions and momentum
distribution are also obtained if the {\it squeezing}
effect and rearrangement of the spin configurations are
taken into account.
In the two-dimensional case, within the mean-field approximation
the magnetized flux state with a
gap in the spinon spectrum has the lowest energy at half-filling. The
antiferromagnetic long-range order is destroyed by hole doping
of the order $\sim 10\div 15$ \%   for $t/J=3\div 5$
and a disordered flux state with  gapless
spinon spectrum becomes stable. The calculated specific heat is roughly
consistent with observed results on copper oxide superconductors.
The possible phase separation is also discussed at the mean-field level.
\end{abstract}


\newpage

\section{Introduction}

The large $U$ Hubbard model and its equivalent, the $t-J$ model
are prototypes to study the strong correlation effects in solids,
especially in connection with the high $T_{c}$
superconductivity \cite{pw1,fcz,yu1}.
The central issue under debate is whether the non-Fermi liquid
behavior, showing up as charge-spin separation and
vanishing of the quasi-particle residue, inherent to the
one-dimensional (1D) Hubbard model is also true for
two-dimensional (2D) models, as conjectured by Anderson \cite{pw2}.

In 1D, the exact Bethe-ansatz solutions \cite{ogata2,ps}
are available for the $t$-$J$ model in the limit
$J/t \rightarrow 0$ and $J/t =2$.
The Hubbard model and the $t$-$J$ model in the small $J$ limit
scale to the Luttinger model \cite{ogata2,sorella1,sorella2}.
Using Lieb-Wu's exact wave function, Ogata and Shiba \cite{ogata2}
have shown the existence of an electron Fermi surface as well
as a singular behavior at $k\sim k_{F}$ and $k\sim 3k_{F}$ in the
electron-momentum distribution function.
Moreover, Yokoyama and Ogata \cite{ogata3},
and Assaad and W\"urtz \cite{ffa} have
studied the 1D $t$-$J$ model using the exact diagonalization of
small systems and quantum Monte Carlo methods, respectively,
and their results show
that the $t$-$J$ model behaves like Luttinger liquids for low
values of $J/t$, and undergoes phase separation at large
values of $J/t$.
Hellberg and Mele \cite{mele} came to the same conclusion
by using  the Jastrow variational wave function.
Thus the typical behavior of the Luttinger liquid \cite{fdm}
in 1D, $i.e.$, the absence of quasi-particle propagation
and charge-spin
separation, has been demonstrated explicitly for the $t$-$J$ model
in the small $J$ limit.

There are no exact solutions available in 2D. The variational
calculations \cite{vg}  seems to support Anderson's conjecture.
The quantum Monte Carlo simulations gave some hint at vanishing of
the quasi-particle residue in the thermodynamic limit \cite{sorella3}.
However, this result is not conclusive because of the " fermion
minus sign" problem in the Monte Carlo technique and contrary
results in exact diagonalization of clusters \cite{wsph} as well
as analytic treatments of the single-hole problem \cite{yulu5}.

The crucial requirement \cite{zje} for the $t$-$J$ model
(and the large
$U$ Hubbard model) is to impose the single occupancy
constraint $\sum_{\sigma}C^{\dagger}_{i\sigma}C_{i\sigma}\leq 1$.
An intuitively
appealing approach to implement this constraint and
the charge-spin separation scheme
is the slave particle formalism \cite{lee1,yu1},
where the electron operator is decomposed as
$C_{i\sigma}$=$a^{\dagger}_{i}f_{i\sigma}$
with $a^{\dagger}_{i}$ as the
slave boson and $f_{i\sigma}$ as  fermion and the local constraint
$\sum_{\sigma}f^{\dagger}_{i\sigma}f_{i\sigma}+a^{\dagger}_{i}a_{i}=1$,
or $vice~~versa$, {\it i.e.}, $a^{\dagger}_{i}$ as
fermion and $f_{i\sigma}$ as
boson. Due to the constraint, these particles are also
coupled by a strong gauge field \cite{lee2}, allowed by this
slave-particle
representation. In the mean field approximation (MFA) the spin
(spinons) and charge (holons) degrees of freedom are
fully separated.
However, there are a number of difficulties in this approach.
First of all, in the slave boson version, the antiferromagnetic
correlation is absent for zero doping, so the ground state energy
in the 2D case is high compared with the numerical estimate for
small clusters, and the Marshall sign rule \cite{wm}
is not obeyed \cite{pw3,tkl}.
Alternatively, in the slave fermion approach, the ground state
is antiferromagnetic for the undoped case and persists until
very high doping ($\sim 60$ \%) \cite{dy}.
The large Fermi surface of spinons, present in the slave
boson approach is absent there. Moreover, if we, following the
common practice, let $f_{i\sigma}$ keep track of the spin,
while $a_{i}$ keep track of the charge, satisfying the
sum rules: $\delta = <a^{\dagger}_{i}a_{i}>$ and
$1 - \delta =\sum_{\sigma}<f^{\dagger}_{i\sigma}f_{i\sigma}>$,
where $\delta$ is the hole doping concentration, we find \cite{feng}
that the sum rule for the physical electron
$\sum_{\sigma}<C^{\dagger}_{i\sigma}C_{i\sigma}>=1-\delta$
is not satisfied for both versions. This expectation
value is $(1-\delta)^{2}$
in the slave fermion representation, and $1-\delta^{2}$ in the
slave boson version. Since the total number of
particles does not depend on the interactions, this
difficulty will persist even beyond the MFA, so long as the
spinon and holon expectation value decoupling is assumed.
Furthermore, we have also shown \cite{feng} that the
overall electron distribution does not have the appropriate Fermi
surface within this scheme even for the 1D case.
These are intrinsic difficulties of this decoupling scheme.

In this paper we develop a new scheme, the fermion-spin
transformation, to implement the local constraint and the
charge-spin separation. In this scheme the charge degree
of freedom is represented by a spinless fermion, while the
spin degree of freedom is represented by a {\it hard-core}
boson in terms of Pauli operators (with a projection operator
to be specified later). Using this representation the local
constraint is satisfied in the decoupling scheme in contrast
with the existing slave particle approach \cite{lee1,yu1},
where the local
constraint is replaced by a global one. As a consequence, the
sum rule for the physical electron is obeyed. Moreover, the
{\it hard-core} bosons can be expressed in terms of spinless
fermions via the Jordan-Wigner transformation in 1D \cite{jordan}
and its
generalization in 2D \cite{mele2}.
This is an efficient calculation scheme
which can provide very good results even at the mean-field
level.

Here we summarize our main results. In 1D we can integrate out
the spinless charge field (holons) and obtain an effective
Hamiltonian for an interacting spinon field which behaves like
a Luttinger liquid. Hence the physical electron, as a convolution
of spinon and holon, also behaves like a Luttinger liquid in
consistency with the exact solution \cite{ogata2}.
Moreover, we obtain a
gapless spectrum for both holons and spinons at the mean-field
level which is not true in the slave fermion approach.
The ground state energy at and away from half-filling is in good
agreement with exact results. By going beyond the
MFA to include the "{\it squeezing}" effect and
rearrangement of the spin configurations due to the hole
presence, we obtain not only correct exponents of
correlation functions
and momentum distribution at the Fermi surface but also a
reasonable global distribution function. In 2D we have
considered various phases at and away from half-filling in the
MFA. The magnetized flux state with a gap
in the spinon spectrum has the lowest energy at half-filling.
The antiferromagnetic long-range order (AFLRO) fades away by
hole doping of the order $10\div 15$\% for $t/J=3\div 5$ in
contrast to the Schwinger boson approach where the AFLRO is
destroyed at 62\% doping \cite{dy}. Beyond the critical
concentration,
a disordered flux phase with gapless spectrum becomes stable.
We have also calculated the specific heat and considered the
phase separation issue. The results are consistent with
experiments and numerical simulations, respectively.

The rest of the paper is organized as follows:
In Sec. II, we explain in detail the  fermion-spin
transformation which is exact in the single occupancy Hilbert
space, if a projection operator is introduced to remove
the extra degrees of freedom. We also estimate the errors
introduced by the MFA. In Sec. III we apply the proposed
scheme to 1D $t$-$J$ model within the MFA.
The main results obtained have been mentioned above.
In Sec. IV we calculate the correlation functions and
momentum distribution by introducing two "string" operators
which take care of the "{\it squeezing}" effect and
rearrangement of the  spin configurations. The exponents
thus obtained
agree with the exact results. The applications to 2D at the
mean-field level are described in Sec. V. Finally, in the
concluding Section we make some further remarks to explain
our current
understanding why this simple transformation works so well and
outline some open problems.

\section{Fermion-Spin Transformation to Implement
 the Charge-Spin Separation}

\subsection{Model, Constraints and Sum Rules}

We start from the $t$-$J$ model which can be written as
\begin{equation}
H = -t\sum_{\langle ij\rangle \sigma}
(C^{\dagger}_{i\sigma}C_{j\sigma} + h.c.)
- \mu \sum_{i\sigma}C^{\dagger}_{i\sigma}C_{i\sigma} +
J\sum_{\langle ij\rangle}{\bf S}_{i}\cdot {\bf S}_{j} ,
\end{equation}
where $C^{\dagger}_{i\sigma}$ ($C_{i\sigma}$) are the electron
creation (annihilation) operators,
${\bf S}_{i}=C^{\dagger}_{i}{\bf \sigma} C_{i}/ 2$
spin operators with
${\bf \sigma}$ as Pauli matrices, $\mu$ the chemical potential.
The summation
$\langle ij\rangle$ is carried over nearest neighbour
nonrepeated bonds.

The Hamiltonian (1) is defined in a restricted Hilbert space
without double electron occupancy. There are two ways to
implement this requirement: either to solve (1) combined
with a nonholonomic constraint
$\sum_{\sigma}C^{\dagger}_{i\sigma}C_{i\sigma}\leq 1$ or
to introduce constrained  fermion operators \cite{rice3},
replacing $C_{i\sigma}$
by $\tilde{C}_{i\sigma}=C_{i\sigma}(1-n_{i\bar{\sigma}})$,
where $n_{i\sigma}=C^{\dagger}_{i\sigma}C_{i\sigma}$. We will
use both representations in this article.

The constrained operators $\tilde{C}_{i\sigma}$ satisfy
the following relations
\begin{equation}
\sum_{\sigma}\tilde{C}^{\dagger}_{i\sigma}\tilde{C}_{i\sigma}
= \sum_{\sigma}C^{\dagger}_{i\sigma}C_{i\sigma}
(1-n_{i\bar{\sigma}}) , ~~~~~~~
\langle \sum_{\sigma}\tilde{C}^{\dagger}_{i\sigma}
\tilde{C}_{i\sigma}\rangle = 1 - \delta  ,
\end{equation}
where the latter equation is a sum rule for the electron at
the hole doping concentration
$\delta$, and $\langle \cdot \cdot \cdot \rangle$
means thermodynamical average. The on-site anticommutation
relation
of the constrained electron operator $\tilde{C}_{i\sigma}$ is
\begin{equation}
\sum_{\sigma}\{\tilde{C}_{i\sigma},\tilde{C}^{\dagger}_{i\sigma}\}
=2- \sum_{\sigma}C^{\dagger}_{i\sigma}C_{i\sigma},~~~~~~~
\langle\sum_{\sigma}\{\tilde{C}_{i\sigma},
\tilde{C}^{\dagger}_{i\sigma}\}
\rangle = 1 + \delta ,
\end{equation}
which gives rise to a sum rule for the  spectral function
$A_{c\sigma}(k,w)$
\begin{equation}
\sum_{\sigma}\int_{-\infty}^{\infty}
{dw \over 2\pi} A_{c\sigma}(k,w)
= 1 + \delta .
\end{equation}
Of course, this value is less than 2 since
$1-\delta$ states are pushed to infinity as
$U\rightarrow \infty$ in deriving the $t$-$J$ model. Hence the
electron spectral function $A_{c\sigma}(k,w)$ describes only
the lower Hubbard band. Eqs. (2) and (4) are  exact
sum rules for the $t$-$J$ model, and they must be
preserved in adequate treatments.

\subsection{CP$^{1}$ Hard-Core Boson}

The decoupling of the charge and spin degrees of freedom for the
physical electron is undoubtedly correct in the 1D $t$-$J$ model
\cite{ogata2}, but the situation is still not clear in 2D. In this
paper, we presume that the decoupling of charge and spin
degrees of freedom for physical electron is also valid for the
2D $t$-$J$ model, and propose a new scheme to decouple the
charge and spin degrees of freedom.

To motivate this transformation,
we start from the no-double occupancy local constraint
$\sum_{\sigma}C^{\dagger}_{i\sigma}C_{i\sigma}\leq 1$. Suppose
$C_{i\sigma}=h^{\dagger}_{i}b_{i\sigma}$ with the spinless
fermion operator
$h_{i}$ keeping track of the charge (holon) while the operator
$b_{i\sigma}$ keeping track of the spin (spinon),
then this on-site local constraint can be rewritten
as $\sum_{\sigma}h_{i}h^{\dagger}_{i}
b^{\dagger}_{i\sigma}b_{i\sigma}\leq 1$. Since the electron
obeys the Fermi statistics, the operator $b_{i\sigma}$ must
be a boson when the operator $h_{i}$ has been assigned a
fermion character in this electron decoupling scheme.
If bosons are subject to the condition
$\sum_{\sigma}b^{\dagger}_{i\sigma}b_{i\sigma}=1$,
the on-site electron local constraint
\begin{equation}
n_{c}=\sum_{\sigma}C^{\dagger}_{i\sigma}
C_{i\sigma}=h_{i}h^{\dagger}_{i}=1-h^{\dagger}_{i}h_{i}\leq 1
\end{equation}
is exactly satisfied, where $n_{h}=h^{\dagger}_{i}h_{i}$
is the spinless holon number at site $i$, equal to  1 or 0.
This way the nonholonomic on-site electron constraint
is converted into
a holonomic boson constraint.

We should note that so long as $h^{\dagger}_{i}h_{i}=1$,
$\sum_{\sigma}C^{\dagger}_{i\sigma}C_{i\sigma}=0$, no
matter what is the value
$n_{b}=\sum_{\sigma}b^{\dagger}_{i\sigma}b_{i\sigma}$.
However, the choice $n_{b}=1$ is convenient, because it
also guarantees the condition $n_{c}=1$, when $n_{h}=0$.
This decoupling scheme, so called CP$^{1}$ representation was
proposed by Wiegmann \cite{pbw} and was used in Ref. 29.
The constraint $n_{c}=1$ means the presence of one boson
(spin-up or -down) on each site, {\it i.e.}, we assign a "spin"
even to an empty site. This will not affect the physical
expectation values, because the hole number expectation
$\langle n_{h} \rangle$ will remove the spurious effects.
Nevertheless, the extra degrees of freedom will affect
the partition function and thermodynamical quantities.
In the next subsection we
will  define a projection operator to cure this defect. As a
result, the commutation relations and sum rules (2)-(4)
will be satisfied exactly.

Now we explore further the properties of the CP$^{1}$
bosons. First note if we restrict the boson occupation
number to 0 or 1,
the infinite-dimensional Fock space for bosons become
two-dimensional, where we can choose the following
representation
\begin{equation}
b_{i}=\left(\matrix{0 &0\cr 1 &0\cr} \right),~~~~
b^{\dagger}_{i}=\left(\matrix{0 &1\cr 0 &0\cr} \right),
\end{equation}
which are nothing but spin-lowering$S^-$ and spin-raising $S^+$
operators for $S=1/2$ and satisfy the {\it hard-core}
constraints $bb=b^{\dagger}b^{\dagger}=0$.

Moreover, if we request that $\uparrow$ and $\downarrow$
{\it hard-core} bosons satisfy the CP$^{1}$ condition
$n_{b}=1$, the $2\times 2$ representation space becomes
two-dimensional. Assume
$\left(\matrix{ 1 \cr 0 \cr} \right )_{\uparrow}$,
$\left(\matrix{ 0 \cr 1 \cr} \right )_{\uparrow}$
are singly occupied and empty spin-up, while
$\left(\matrix{ 0 \cr 1 \cr} \right )_{\downarrow}$,
$\left(\matrix{ 1 \cr 0 \cr} \right )_{\downarrow}$
are singly occupied and empty spin-down states, respectively.
Due to the constraint
$b^{\dagger}_{i\uparrow}b_{i\uparrow}+b^{\dagger}_{i\downarrow}
b_{i\downarrow}=1$, out of 4 possible states as direct
products, only two, namely
$\left(\matrix{ 1 \cr 0 \cr} \right )_{\uparrow}$
$\left(\matrix{ 1 \cr 0 \cr} \right )_{\downarrow}$, and
$\left(\matrix{ 0 \cr 1 \cr} \right )_{\uparrow}$
$\left(\matrix{ 0 \cr 1 \cr} \right )_{\downarrow}$
are allowed. Thus we can ignore the spin label in the state and
represent $b_{\uparrow}$ as
$\left(\matrix{0 &0\cr 1 &0\cr} \right)$, and $b_{\downarrow}$
as $\left(\matrix{0 &1\cr 0 &0\cr} \right)$ in the reduced
two-dimensional representation space. Of course,
all the {\it hard-core boson} conditions, {\it i.e.},
$b_{i\sigma}b^{\dagger}_{i\sigma}+b^{\dagger}_{i\sigma}b_{i\sigma}=1$,
$b^{\dagger}_{i\sigma}b^{\dagger}_{i\sigma}=b_{i\sigma}b_{i\sigma}=0$,
(without summation over $\sigma$), are satisfied. As a result,
$b_{\uparrow}$ and $b_{\downarrow}$ are identified with the
spin lowering
$S^{-}$ and raising $S^{+}$ operators, respectively, while the
boson occupation space is identified with the spin 1/2 representation
space.

To sum up, as solutions of the single occupancy constraint
$\sum_{\sigma}C^{\dagger}_{i\sigma}C_{i\sigma}\leq 1$ under CP$^{1}$
convention $\sum_{\sigma}b^{\dagger}_{i\sigma}b_{i\sigma}=1$,
we find the following fermion-spin transformation
\begin{equation}
C_{i\uparrow}=h^{\dagger}_{i}S^{-}_{i},~~~~
C_{i\downarrow}=h^{\dagger}_{i}S^{+}_{i}
\end{equation}
in terms of which the $t$-$J$ Hamiltonian (1) can be rewritten as
\begin{eqnarray}
H = -t\sum_{\langle ij\rangle\sigma}h_{i}h^{\dagger}_{j}
(S^{+}_{i}S^{-}_{j} + S^{-}_{i}S^{+}_{j}) + h.c. \nonumber \\
 - \mu \sum_{i}h_{i}h^{\dagger}_{i}
+ J\sum_{\langle ij\rangle}
(1-h^{\dagger}_{i}h_{i}){\bf S}_{i}\cdot {\bf S}_{j}
(1-h^{\dagger}_{j}h_{j}) ,
\end{eqnarray}
{\it where ${\bf S}_{i}$ is the pseudo-spin operator} at site $i$
which can be
expressed as $CP^{1}$ {\it hard-core bosons} and is
different from
{\it the electron spin operator} in Eq. (1). We would emphasize
that the  present CP$^{1}$ {\it hard-core boson}
representation of spin
operators is different from the CP$^{1}$ boson representation
of spin operator used before \cite{pbw,weng}. In their approach,
the spin
degree of freedom is represented by ordinary boson operators,
while, in the
present scheme, the {\it hard-core} boson operator $b_{i\sigma}$
behaves as
a fermion on the same site, and as a  boson on different sites.

\subsection{Projection Operator}

In the local representation  the restricted Hilbert space
of no-double occupancy consists of three
states, $|0\rangle$, $|\uparrow \rangle$,
$|\downarrow \rangle$, while in the fermion-spin transformation
presented in the previous Subsection there are four states
$|hole\rangle \otimes |spin \rangle$, namely
$|1,\uparrow \rangle$, $|1,\downarrow \rangle$, $|0,\uparrow \rangle$,
and $|0,\downarrow \rangle$, where 1 or 0 means hole occupation.
We can introduce a projection operator $P$ to remove the extra
degrees of freedom. The matrix elements of this operator can be defined as
\begin{equation}
P_{\kappa \alpha}\equiv |\kappa \rangle \langle \alpha |,
\end{equation}
where $|\kappa \rangle$ is one of the bases of the physical states,
while $|\alpha \rangle$ is one of the bases in the space
$|hole \rangle \otimes |spin \rangle$. Since the space dimensions of
 $|\kappa \rangle$  and  $|\alpha \rangle$  are different,
the usual relations
for the projection operator $P^2=P=P^{\dagger}$ are
not satisfied.  Using this operator,
 one can define the electron operators in the constrained space as
\begin{eqnarray}
\tilde{C}_{i\uparrow}=P_{i}h^{\dagger}_{i}S^{-}_{i}P^{\dagger}_{i},~~~
\tilde{C}_{i\downarrow}=P_{i}h^{\dagger}_{i}S^{+}_{i}P^{\dagger}_{i},
\nonumber \\
\tilde{C}^{\dagger}_{i\uparrow}=P_{i}h_{i}S^{+}_{i}P^{\dagger}_{i},~~~
\tilde{C}^{\dagger}_{i\downarrow}=P_{i}h_{i}S^{-}_{i}P^{\dagger}_{i},
\end{eqnarray}
where $P_i$ is the projection operator for the site $i$,
 $P_i^{\dagger}$ is the
hermitian conjugate of $P_i$. Making use of the matrix
representation of the
holon operator $h^{\dagger}_{i}=\left(\matrix{0 &1\cr 0 &0\cr} \right)$,
$h_{i}=\left(\matrix{0 &0\cr 1 &0\cr} \right)$,
we can write down explicitly all these operators in matrix form
(see Appendix). In particular, the constrained electron operators
in the basis of the physical states $|0\rangle$, $|\uparrow \rangle$,
$|\downarrow \rangle$ can be written as
\begin{eqnarray}
\tilde{C}_{i\uparrow}=
\left(\matrix{0&1&0\cr 0&0&0\cr 0&0&0\cr} \right),~~~~
\tilde{C}^{\dagger}_{i\uparrow}=
\left(\matrix{0&0&0\cr 1&0&0\cr 0&0&0\cr} \right),\nonumber \\
\tilde{C}_{i\downarrow}=
\left(\matrix{0&0&1\cr 0&0&0\cr 0&0&0\cr} \right),~~~~
\tilde{C}^{\dagger}_{i\downarrow}=
\left(\matrix{0&0&0\cr 0&0&0\cr 1&0&0\cr} \right),
\end{eqnarray}
which are nothing but the Hubbard $X$ operators
$X_{0\uparrow}$, $X_{0\downarrow}$, etc. \cite{hubbard}. It is then
straightforward to check that
\begin{equation}
\sum_{\sigma}\tilde{C}^{\dagger}_{i\sigma}\tilde{C}_{i\sigma}
= 1-n_{h},
\end{equation}
\begin{equation}
\sum_{\sigma}\{\tilde{C}_{i\sigma},\tilde{C}^{\dagger}_{i\sigma}\}
=1+n_{h} ,
\end{equation}
where $n_{h}=\left(\matrix{1&0&0\cr 0&0&0\cr 0&0&0\cr} \right)$
is the hole number operator.
Taking expectation values of Eqs. (12) and (13), one sees
immediately that the sum rules  (2) and (4) are exactly satisfied.
Thus we have shown that the fermion-spin transformation defined
with an additional projection operator $P$  satisfies exactly
the no-double occupancy constraint and all sum rules,$i.e.$, they
are an exact mapping.

However, the projection operator $P$ is cumbersome to handle
and in many cases, for example in the mean-field treatment of
Sections III-V, we will drop it. Now let us see which of these
properties are still preserved and what kind of errors we are
committing in such approximate treatments. First of all, the
local constraints are exactly obeyed  even in the MFA. Secondly,
those expectation values of electron operators, including spin-spin
correlation functions, which should vanish, actually do not
appear due to the presence of the holon number operator
$n_{h}=h^{\dagger}_{i}h_{i}$. Furthermore, as we will see later,
the sum rules  for the physical electron are also satisfied in
MFA. By adding an extra spin degree of freedom to an empty
site we are making errors of the order $\delta$ in counting
the number of states, which is negligible if
$\delta \rightarrow 0$.
For comparison we should note that in the usual
slave-particle approach \cite{lee1},
the local constraint is
explicitly replaced by a global constraint in MFA,
and therefore the representation space is much
larger than the  representation space for the physical electron,
which leads to some unphysical results \cite{zje,feng}.
 From this point of view, our
treatment of  constraints for the physical electron is much better
than the slave-particle approach, and therefore we believe the
mean-field result based on the fermion-spin approach even without
projection operator should be better than those obtained by
using the slave-particle mean-field theory. This is indeed
confirmed by the mean-field calculations presented in the
following Sections.

We note that a similar transformation has been discussed in
Ref. 31, but these authors did not stick to the
single-occupancy constraints in their actual calculations.

\section{The Mean Field Theory in 1D
within the Fermion-Spin Approach }

Since an exact solution \cite{wu} for the Hubbard model (hence
for the $J/t \rightarrow 0$ limit of the $t$-$J$ model) is
available in 1D, it is important to confront any approximate
treatment with this solution. In this Section we consider the
1D $t$-$J$ model, using the fermion-spin transformation described
in the previous Section at the mean-field level, neglecting the
effects of the projection operator $P$. As mentioned in Introduction,
the 1D $t$-$J$ model exhibits a Luttinger liquid behavior, including
charge-spin separation, vanishing of the quasi-particle-residue,
etc. We should also mention that both spin and charge excitations
are gapless in 1D \cite{ogata2}. In the standard slave-particle
approach, many of these properties are not preserved. For example,
in the slave-boson case \cite{zwang} there is a Bose condensation
at the mean-field level which leads to a Fermi-liquid
behavior. On the other hand, there is a gap in the spin
excitation spectrum in 1D within the slave-fermion framework,
even at half-filling \cite{dy}. Now consider the mean-field
results in the fermion-spin approach.

\subsection{Luttinger liquid behavior}

In the fermion-spin representation, the 1D $t$-$J$ model
may be written as
\begin{eqnarray}
H = -t\sum_{i}a^{\dagger}_{i}a_{i+1}(S^{+}_{i}S^{-}_{i+1} +
S^{-}_{i}S^{+}_{i+1}) +  h. c.  \nonumber \\
- \mu \sum_{i}a^{\dagger}_{i}a_{i} +
J\sum_{\langle ij\rangle}(a^{\dagger}_{i}a_{i})({\bf S}_{i}
\cdot {\bf S}_{j})(a^{\dagger}_{j}a_{j}) ,
\end{eqnarray}
where, for the convenience of the following discussion, we have
introduced the particle operators $a_{i}$ and $a^{\dagger}_{i}$
defined as
\begin{equation}
a_{i}=h^{\dagger}_{i}, ~~~~~~~ a^{\dagger}_{i}=h_{i} .
\end{equation}
In the 1D  case, the {\it hard-core bosons}
$S^{+}_{i}$ and $S^{-}_{i}$ can be mapped exactly onto spinless
fermions using the Jordan-Wigner \cite{jordan} transformation
\begin{equation}
S^{+}_{i} = f^{\dagger}_{i}e^{i\pi \sum_{l<i}f^{\dagger}_{l}f_{l}} ,
\end{equation}
\begin{equation}
S^{-}_{i} = f_{i}e^{-i\pi \sum_{l<i}f^{\dagger}_{l}f_{l}} ,
\end{equation}
\begin{equation}
S^{Z}_{i} = f^{\dagger}_{i}f_{i} - {1\over 2} ,
\end{equation}
where $f_{i}$ is the spinless fermion operator.
Substituting Eqs. (16)-(18) into Eq. (14), the $t$-$J$
Hamiltonian (14) can be expressed as
\begin{eqnarray}
H = -t\sum_{i}(a^{\dagger}_{i}a_{i+1}+a^{\dagger}_{i+1}a_{i})
(f^{\dagger}_{i}
f_{i+1}+f^{\dagger}_{i+1}f_{i}) -\mu \sum_{i}a^{\dagger}_{i}a_{i}
\nonumber \\
+J\sum_{\langle ij\rangle}a^{\dagger}_{i}a_{i}[{1\over 2}
(f^{\dagger}_{i}f_{j}
+ f^{\dagger}_{j} f_{i}) + (f^{\dagger}_{i}f_{i}-{1\over 2})
(f^{\dagger}_{j}f_{j}-{1\over 2})]a^{\dagger}_{j}a_{j} .
\end{eqnarray}
We can now employ the path-integral representation in which the
Lagrangian $L$ and the partition function $Z$ of the
$t$-$J$ model in the imaginary time $\tau$ can be expressed as
\begin{equation}
L = \sum_{i}a^{\dagger}_{i}\partial_{\tau}a_{i} + \sum_{i}f^{\dagger}_{i}
\partial_{\tau}f_{i} + H ,
\end{equation}
\begin{equation}
Z = \int DaDa^{\dagger}DfDf^{\dagger}e^{-\int d\tau L(\tau)} .
\end{equation}
Integrating out the spinless charge field $a_{i}$ of the
$t$-$J$ model, we obtain an interacting spinon system, which
like any interacting fermion systems in 1D,
is described by a  Luttinger liquid theory \cite{fdm,luther}.
Moreover, the physical electron as a convolution of spinon and
holon also behaves like a
Luttinger liquid, which means that the electron wave function
renormalization constant $Z=0$ in the 1D $t$-$J$ model.

In the path-integral representation, one can introduce a
$SU(2)$-invariant Hubbard-Stratonovich transformation
to decouple the Lagrangian (20) by using the
following auxiliary fields
\begin{equation}
\chi_{i,i+\eta} = S^{+}_{i}S^{-}_{i+\eta} ,
\end{equation}
\begin{equation}
\phi_{i,i+\eta} = a^{\dagger}_{i}a_{i+\eta} ,
\end{equation}
where $\eta = \pm 1$. Note the auxiliary field
$\chi_{i,i+\eta}$ is a boson type field. MFA to the
$t$-$J$ model (19) is just the saddle point
solution of the Lagrangian (20),
{\it i.e.}, the auxiliary fields
$\chi_{i,i+\eta}$ and $\phi_{i,i+\eta}$ are replaced
by their mean-field values
$\chi_{i,i+\eta}=\chi$ and $\phi_{i,i+\eta}=\phi$,
respectively,
and the Hamiltonian (19) can be diagonalized as
\begin{equation}
H = \sum_{k}\varepsilon (k)a^{\dagger}_{k}a_{k} +
\sum_{k}\omega (k)f^{\dagger}_{k}f_{k} + 4Nt\chi \phi +
2NJ[(1-\delta)^{2}-\phi^{2}]\chi^{2} ,
\end{equation}
where N is the number of sites, and
\begin{equation}
\varepsilon (k) =-4t\chi {\rm cos}(k) -\mu ,
\end{equation}
\begin{equation}
\omega (k) = [((1-\delta)^{2}-\phi^{2})(1-2\chi )-{2t\over J}
\phi ](2J){\rm cos}(k),
\end{equation}
while the self-consistent equations for the order parameters
$\chi$ and $\phi$ can be obtained by minimizing the free energy.
We can now proceed to a brief discussion of the results in MFA.

\subsection{Groud-State Properties at Half-Filling}

At half-filling, there are no charge degrees of freedom,
and the $t$-$J$ model (24) reduces to the antiferromagnetic
Heisenberg model in the fermion representation
\begin{equation}
H_{J} = \sum_{k}\omega_{0}(k)f^{\dagger}_{k}f_{k} + 2NJ\chi^{2} .
\end{equation}
In this case the order parameter $\chi$ can be evaluated
to be $\chi = - 1/\pi$,
and we obtain a gapless spinon spectrum,
\begin{equation}
\omega_{0}(k) = (1+{2\over \pi})(2J){\rm cos}(k),
\end{equation}
which is rather close to
the exact result of the Heisenberg chain obtained
by using the Bethe-ansatz method \cite{bethe}
$\omega_{0}(k)=(\pi /2)(2J){\rm cos}(k)$.
Correspondingly, the spinon spectrum near the spinon
Fermi surface ($k=\pm \pi /2$) is linear with velocity
$v_{s}=(1+2/\pi)(2J) =1.6366(2J)$, which is also very close to the
exact result of the Heisenberg chain obtained by Haldane \cite{fdmh}
$v_{s}=(\pi /2)(2J)=1.5708(2J)$.
The ground state energy of the Heisenberg model
at temperature $T=0$ is
$E_{0} = -0.4196(2J)$
which is only 5.3 percent higher than the exact Bethe-ansatz
value \cite{bethe} of $E_{0} =-( 1/4 - {\rm ln}2 )(2J) = -0.4431(2J)$.

Thus in the half-filled case, the spinon has
a gapless spectrum, and the spinon ground state
energy can be described adequately by the Jordan-Wigner
transformation within MFA. This case has already been considered earlier
\cite{weng,wang}.

\subsection{Ground-State Energy away from Half-Filling}

Away from half-filling, a gapless holon spectrum is obtained
\begin{equation}
\varepsilon (k) = -2t{2\over \pi}{\rm cos}(k) ,
\end{equation}
from which the holon Fermi velocity is given by
\begin{equation}
v_{h} = 2t{2\over \pi}{\rm sin}[(1-\delta)\pi] = 2t{2\over \pi}
{\rm sin}(\delta \pi ),
\end{equation}
and the holon ground state energy at temperature $T=0$
is obtained as
\begin{equation}
E_{h} = -({2t\over \pi})
({2\over \pi}){\rm sin}(\delta \pi ) .
\end{equation}
All these quantities differ from the corresponding exact values
\cite{ogata2} of the 1D
large $U$ limit Hubbard model
$\varepsilon (k) = -2t {\rm cos}(k)$,
$v_{h}=2t{\rm sin}(\delta \pi)$, and
$E_{h}=-({2t\over \pi}){\rm sin}(\delta \pi)$, only by
a factor 2/$\pi$.

At the same time, the gapless spinon spectrum at finite
dopings becomes
\begin{equation}
\omega (k) =[((1-\delta )^{2}-{{\rm sin}^{2}(\delta \pi)
\over \pi^{2}})(1+{2\over \pi}) + {2t\over J}
{1\over \pi}{\rm sin}(\delta \pi )](2J){\rm cos}(k) ,
\end{equation}
with the spinon ground state energy
\begin{equation}
E_{s} = -(1-\delta)^{2}[1-{{\rm sin}^{2}(\delta \pi)\over \pi^{2}
(1-\delta)^{2}}]E_{0} ,
\end{equation}
which is again very close to the exact result of
the spinon ground state energy \cite{ogata2} of
the 1D large $U$ limit Hubbard model over the
entire doping range. Here $E_{0}$ is the ground state energy at
half-filling. Therefore the total ground state energy of
the 1D $t$-$J$ model in the fermion-spin approach within MFA
can be expressed as
\begin{eqnarray}
E_{g} = E_{h} + E_{s} ~~~~~~~~~~~~~~~~~~~~~~~~~~~~~~~~~~~~~~~~~~~~~~~~
{}~~~~~ \nonumber \\
= -({2t\over \pi })({2\over \pi }){\rm sin}(\delta \pi ) -
(1-\delta )^{2}[1 - {{\rm sin}^{2}(\delta \pi )\over \pi^{2}
(1-\delta )^{2}}]
0.4196(2J) .
\end{eqnarray}
The ground state energy and thermodynamical quantities for the
1D Hubbard model
in the atomic limit have been calculated long time back\cite{klein}.
Our result for the ground state energy is rather close to theirs.

\subsection{Momentum Distribution}

Now we turn to discuss the global features of the electron
momentum distribution within MFA. This
distribution for physical electron is defined as
\begin{equation}
n(k)=\sum_{\sigma}\langle \tilde{C}^{\dagger}_{k\sigma}
\tilde{C}_{k\sigma}\rangle=
{1\over N}\sum_{lm\sigma}e^{ik(x_{l}-x_{m})}
\langle \tilde{C}^{\dagger}_{l\sigma}\tilde{C}_{m\sigma}\rangle .
\end{equation}
In the present fermion-spin approach, neglecting the effects due
to the projection operator defined in Sec. II C, this distribution
function can be rewritten as
\begin{equation}
n(k)={1\over N}\sum_{lm}e^{ik(x_{l}-x_{m})}\langle a^{\dagger}_{l}
a_{m}(S^{+}_{l}S^{-}_{m}+S^{-}_{l}S^{+}_{m})\rangle .
\end{equation}
Using the Jordan-Wigner transformation (16)-(18), $n(k)$ can be
further expressed as
\begin{eqnarray}
n(k)={1\over N}\sum_{lm}e^{ik(x_{l}-x_{m})}\langle a^{\dagger}_{l}
a_{m}(f^{\dagger}_{l}e^{i\pi(\sum_{i<l}f^{\dagger}_{i}f_{i}-
\sum_{j<m}f^{\dagger}_{j}f_{j})}f_{m} \nonumber \\
+ f_{l}e^{-i\pi(\sum_{i<l}f^{\dagger}_{i}f_{i}-\sum_{j<m}
f^{\dagger}_{j}f_{j})}f^{\dagger}_{m})\rangle .
\end{eqnarray}
Since $e^{\pm i\pi f^{\dagger}_{i}f_{i}}=1-2f^{\dagger}_{i}f_{i}$,
in MFA, we obtain
\begin{equation}
n(k)=1-\delta +\sum^{\infty}_{m=1}(-1)^{m+1}({2\over \pi})^{m+1}
{{\rm sin}(m\delta \pi )\over m}{\rm cos}(mk) ,
\end{equation}
which is plotted (solid line) in figure 1 for
doping $\delta =0.5$.
For comparison, the corresponding curves for the
slave-boson (dashed line)
and the slave-fermion (dot-dashed line) cases are also given. It
is obvious that $n_{k} \geq 1-\delta$ in
the slave-boson case and $\leq 1-\delta$ in the slave-fermion case,
which is very far from a should-be electron momentum distribution.
The integrated area under the curve is equal to
($1-\delta^{2}$) in the slave-boson case and is
$(1-\delta )^{2}$ in the slave-fermion case, while the correct
value should be $1-\delta$ \cite{feng}.
The solid curve corresponding to our
transformation, is closer to the exact result \cite{ogata2}.
The integrated area is correct and the shape looks like
a reasonable momentum distribution, {\it i.e.}, in some part the
distribution is greater, while in other part it is less than
$1-\delta$. To get a more accurate result (including the correct
location of the Fermi surface and a correct slope at it) one
should go beyond the MFA to include the spinon-holon interactions
as discussed in the next Section.

Thus away from  half-filling, the spinons and holons are
decoupled completely, with the holon behaving
like a spinless fermion, while the spinon has the Jordan-Wigner
\cite{jordan} form in 1D. The gapless spectra for both holons
and spinons, as well as  the ground state energy can be
described adequately within the fermion-spin approach even in MFA.

\section{The Asymptotic Behavior of Correlation
functions in the 1D case}

\subsection{Motivation}

Interacting 1D electron systems generally behave like
Luttinger liquids \cite{fdm} where the electron
correlation functions show a power-law decay with unusual
exponents. These systems exhibit
an electron Fermi surface with a correct Luttinger volume but
the momentum distribution function is singular at the Fermi surface,
also with unusual exponents \cite{ogata2}.
These exponents depend on the interaction
strength. Haldane \cite{fdm} has shown that the characteristics of
Luttinger liquids can be calculated using the bosonization techniques.
To get more insight into the problem let us consider the Bethe
ansatz wave function for the Hubbard model in the large $U$
limit (hence for the $t$-$J$ model in the small $J$ limit),
derived by Ogata and Shiba \cite{ogata2}
\begin{equation}
\Psi (x_{1},\cdot \cdot \cdot , x_{N}) =det[exp(ik_{i}x_{Q_{i}})]
\Phi(y_{1}, \cdot \cdot \cdot , y_{M}),
\end{equation}
where the determinant depends only on the coordinates
$x_{Q_{j}}$ of particles, but not on their spins, while
$\Phi(y_{1}, \cdot \cdot \cdot , y_{M})$ is the  Bethe ansatz
wave function for an AF Heisenberg chain with
$y_{1}, \cdot \cdot \cdot , y_{M}$ as coordinates of down
spins \cite{bethe}. This asymptotic form can be interpreted as
a complete  separation of charge and spin degrees of freedom
in some sense. In fact, the determinant describes spinless
fermions (holons), whereas $\Phi$ is the "spinon" wave
function. Our fermion-spin transformation is, to some
extent, an approximate second quantized version of this solution,
with holon being represented by a spinless fermion and spin
represented by a {\it hard-core} boson.

However, there is an important "detail" in the wave function
(39), namely, the spin wave function $\Phi$ is for a
"{\it squeezed}" Heisenberg chain, {\it i.e.}, all holes
should be removed. This will lead to rearrangement of
spin configurations and far-reaching nonlocal effects.
Therefore, spinons and holons are not completely independent,
but interacting with each other rather strongly. As
shown in Ref. 5, the correct exponents of correlation
functions and an adequate description of the momentum
distribution function (Fermi surface at $k_{F}$, rather than
$2k_{F}$ with appropriate singular behavior) can be obtained
only if these interaction effects are properly taken into
account. Weng $et~al$. \cite{weng} have shown that the effects
due to squeezing and rearrangement of spin configurations
can be included by introducing a nonlocal "string" field.
After doing that correct results for the correlation
function exponents, {\it etc.}, can be obtained for a 1D $t$-$J$
chain. In this Section we calculate these exponents within
our fermion-spin approach, following their technique
with some modifications. The nonlocal effects due to
spinon-holon and spinon-spinon interactions will be included
by introducing two string fields.
After squeezing, there are no
holes in the chain, so the additional degrees of freedom
due to assigning "spin" to a hole site disappear, hence
the projection operator introduced in Sec. II C is not needed
for our purpose.

\subsection{String Operators}

Let us consider the  the $t$-term in the $t$-$J$ Hamiltonian
within the fermion-spin representation.
Following Ref. 29, the largest holon kinetic energy may
be obtained if the spin configurations are squeezed as
\begin{equation}
P_{i}(a^{\dagger}_{i}S^{+}_{i}P^{\dagger}_{i}P_{i+1}
a_{i+1}S^{-}_{i+1}
+a^{\dagger}_{i}S^{-}_{i}P^{\dagger}_{i}P_{i+1}a_{i+1}
S^{+}_{i+1})
P^{\dagger}_{i+1} = a^{\dagger}_{i}a_{i+1}
\end{equation}
for all sites $i$ where a holon is present. However, after such
squeezing, the spin configurations are not optimal to favor
the spinon energy. Thus at
the same time spin configurations must be rearranged into optimum
configurations to provide the lowest spinon energy. These optimum
spin configurations can be obtained by reversing the original spin
polarization for all sites on the left-hand side of
site $i$. These processes of
first squeezing the $t-J$ chain and then rearranging the
spin configurations, are shown schematically in figure 2.
These nonlocal processes cannot be described
by any formal perturbation theory, but can be taken into account
approximately by introducing the string fields \cite{weng,feng}.
In our case,
we introduce two string fields to describe the above
physical processes, so the constrained electron operator
$\tilde{C}_{i\sigma}$ can be expressed within the
fermion-spin approach as
\begin{equation}
\tilde{C}_{i\uparrow} = (e^{i\pi (N-\sum_{l>i}a^{\dagger}_{l}
a_{l})}a_{i})(e^{-i\pi \sum_{l<i}a^{\dagger}_{l}a_{l}}S^{-}_{i}) ,
\end{equation}
\begin{equation}
\tilde{C}_{i\downarrow} = (e^{i\pi (N+\sum_{l>i}
a^{\dagger}_{l}a_{l})}a_{i})(e^{i\pi
\sum_{l<i}a^{\dagger}_{l}a_{l}}S^{+}_{i}) ,
\end{equation}
which means that the spinless fermion $a_{i}$ and the
hard-core boson
$S^{\pm}_{i}$ are replaced by corresponding string
operators as
\begin{equation}
a_{i}\rightarrow a_{i}e^{i\pi (N\pm \sum_{l>i}a^{\dagger}_{l}a_{l})} ,
\end{equation}
\begin{equation}
S^{\pm}_{i}\rightarrow S^{\pm}_{i}
e^{\pm i\pi \sum_{l<i}a^{\dagger}_{l}a_{l}} ,
\end{equation}
where $e^{\pm i\pi \sum_{l<i}a^{\dagger}_{l}a_{l}}$ is the
string field for the spinon due to the presence of holons,
describing the effects of rearrangements of spin configurations from
-$\infty$ to site $i$ when an electron is removed or added at site $i$.
For the $t-J$ chain, holons on the right-hand side of site $i$ can
feel some indirect effects of holons on the left-hand side of site
$i$ due to the rearrangements of spin configurations from -$\infty$ to
site $i$ when one electron was removed or added at site $i$. These
indirect effects can be described by the string field
$e^{i\pi (N\pm \sum_{l>i}a^{\dagger}_{l}a_{l})}$  for
the holon. One can check easily that the anticommutation
relations for the physical electron are preserved exactly in
our case.

\subsection{Energy Spectrum for a Squeezed Chain}

After {\it squeezing} and rearranging the spin configurations, the
$t$-$J$ model can be written as
\begin{equation}
H=-t\sum_{i}(a^{\dagger}_{i}a_{i+1}+h. c. ) -
\mu \sum_{i}a^{\dagger}_{i}a_{i}
+ J[(1-\delta)^{2}-\phi^{2}]\sum_{\langle ij\rangle}
{\bf S}_{i}\cdot {\bf S}_{j},
\end{equation}
where we have approximated the probability of the spin exchange
process of the Heisenberg term
as $a^{\dagger}_{i}a_{i} a^{\dagger}_{j}a_{j}\approx
\langle a^{\dagger}_{i}a_{i}\rangle
\langle a^{\dagger}_{j}a_{j}\rangle -\langle a^{\dagger}_{i}
a_{j}\rangle \langle a^{\dagger}_{j}a_{i}\rangle
=(1-\delta )^{2}-\phi^{2}$, and the lattice
constant of the {squeezed} spin chain has become $a/(1-\delta )$,
where $a$ is the original
lattice constant. Therefore the
Fermi points of spinons are shifted
from $k_{F}=\pm \pi/2$ of the half-filled case to
$k_{F}=\pm \pi (1-\delta)/2$ for the doped case.

For the {\it squeezed} chain, the gapless holon spectrum,
the holon Fermi velocity, and the holon
ground state energy are
\begin{equation}
\varepsilon (k) = -2t{\rm cos}(k) ,
\end{equation}
\begin{equation}
v_{h}=2t{\rm sin}(\delta \pi),
\end{equation}
\begin{equation}
E_{t} = -{2t \over \pi}{\rm sin}(\delta \pi) ,
\end{equation}
respectively, which are in full agreement with the
corresponding exact values of the 1D large $U$  Hubbard
model \cite{ogata2}.
At the same time, the gapless spinon spectrum, spinon
Fermi velocity, and the
spinon ground state energy are given by
\begin{equation}
\omega (k) = [(1-\delta )^{2}-{{\rm sin}^{2}(\delta \pi )\over
\pi^{2}}](2J){\rm cos}(k) ,
\end{equation}
\begin{equation}
v_{s}=[(1-\delta )^{2}-{\rm sin}^{2}(\delta \pi )/ \pi^{2}](2J)
{\rm cos}(\delta \pi /2),
\end{equation}
\begin{equation}
E_{J} = [(1-\delta)^{2} - {{\rm sin}^{2}(\delta \pi )\over \pi^{2}}]E_{0} ,
\end{equation}
respectively, which are very close to the corresponding exact
values of the 1D large $U$ limit Hubbard model \cite{ogata2}.
Here $E_{0}$ is the ground state energy at half-filling.
 Therefore, the total energy of the $t$-$J$
model $E_{g}=E_{t}+E_{J}$  agrees quantitatively with
the exact value of the
1D large $U$ limit Hubbard model in the entire doping range. It fully
agrees with the 1D results for the Hubbard model obtained
earlier in the atomic limit \cite{klein}. The gapless
spinon and holon spectra, and the ground state energy are all better than
the mean-field results obtained in Section III, which indicates that
the string fields take into account the spinon-holon interactions
in the $t$-$J$ model, and renormalize
considerably the results obtained in MFA without string fields.

\subsection{Correlation Functions}

The spin-spin correlation function is defined as
\begin{eqnarray}
S(x_{i}-x_{j},t) = \langle S^{X}_{i}(t)S^{X}_{j}(0)\rangle
=\langle S^{Y}_{i}(t)S^{Y}_{j}(0)\rangle
\nonumber \\
=\langle S^{Z}_{i}(t)S^{Z}_{j}(0)\rangle
={1\over 4}\langle S^{+}_{i}(t)S^{-}_{j}(0) +
S^{-}_{i}(t)S^{+}_{j}(0)\rangle .
\end{eqnarray}
In the doped case, we need to replace the operator $S^{\pm}_{i}$
in Eq. (52) by Eq. (44) to account for
the presence of holons. Thus substituting Eq. (44) into
Eq. (52), we
obtain the spin-spin correlation function of the $t$-$J$ model as
\begin{eqnarray}
S(x_{i}-x_{j}, t) ={1\over 4}[\langle S^{+}_{i}(t)
S^{-}_{j}(0)\rangle
\langle e^{i\pi \sum_{l<i}a^{\dagger}_{l}(t)a_{l}(t)}
e^{-i\pi \sum_{l<j}
a^{\dagger}_{l}(0)a_{l}(0)}\rangle \nonumber \\
+\langle S^{-}_{i}(t)S^{+}_{j}(0)\rangle
\langle e^{-i\pi \sum_{l<i}a^{\dagger}_{l}(t)a_{l}(t)}
e^{i\pi \sum_{l<j}a^{\dagger}_{l}(0)a_{l}(0)}\rangle ] .
\end{eqnarray}
Following Haldane \cite{fdm,weng}, we apply
the bosonization method to the
free holon field, and obtain the following asymptotic form
\begin{equation}
\langle e^{\pm i\pi \sum_{l<i}a^{\dagger}_{l}(t)
a_{l}(t)}e^{\mp i\pi \sum_{l<j}
a^{\dagger}_{l}(0)a_{l}(0)}\rangle \sim {1\over
[(x_{i}-x_{j})^{2} - (v_{h}t)^{2}]^{{1\over 4}}} ,
\end{equation}
where $v_{h}$ is the holon Fermi velocity. Luther and
Peschel \cite{luther} have mapped the 1D Heisenberg model into
an interacting  spinless fermion system by using
the Jordan-Wigner \cite{jordan} transformation. Their work involves
generalization of the Jordan-Wigner transformation to provide a
representation for continuum spin operators.
The asymptotic form of the spin-spin correlation
function of the Heisenberg model can be then
obtained by considering the
spinon-spinon interactions. Following their calculation, we get
\begin{equation}
\langle S^{+}_{i}(t)S^{-}_{j}(0)+S^{-}_{i}(t)
S^{+}_{j}(0)\rangle \sim
{{\rm cos}[2k_{F}(x_{i}-x_{j})]
\over [(x_{i}-x_{j})^{2}-(v_{s}t)^{2}]^{{1\over 2}}} ,
\end{equation}
where $v_{s}$ is the spinon Fermi velocity. Combining Eq. (54)
with Eq. (55), we obtain the asymptotic form of the
spin-spin correlation function of the 1D $t$-$J$  model as
\begin{equation}
S(x_{i}-x_{j}, t) \sim
{1\over [(x_{i}-x_{j})^{2}-(v_{h}t)^{2}]^{{1\over 4}}}
\cdot {{\rm cos}[2k_{F}(x_{i}-x_{j})]
\over [(x_{i}-x_{j})^{2}-(v_{s}t)^{2}]^{{1\over 2}}} ,
\end{equation}
which shows a power-law decay with exponent 3/2,
in full agreememt with the numerical result
of the 1D large $U$ limit Hubbard model obtained
by Ogata and Shiba \cite{ogata2}.

\subsection{Momentum Distribution Function}

The electron momentum distribution function $n(k)$ is defined
in Eq. (36)
within the fermion-spin approach. To consider the {\it squeezing}
effect and rearrangements of the spin configurations, we need to
substitute Eqs. (43)-(44) into Eq. (36). Then $n(k)$ can
be rewritten as
\begin{equation}
n(k)={2\over N}\sum_{ij}e^{ik(x_{i}-x_{j})}\langle
e^{i{\pi \over 2}\sum_{l<i}
a^{\dagger}_{l}a_{l}}a^{\dagger}_{i}S^{+}_{i}
G^{*}_{i}G_{j}S^{-}_{j}a_{j}
e^{-i{\pi \over 2}\sum_{l<j}a^{\dagger}_{l}a_{l}}\rangle ,
\end{equation}
where the factor
$G_{i}=e^{i{\pi \over 2}(N-\sum_{l>i}a^{\dagger}_{l}a_{l})}$.
In what follows, we will drop this factor as in the
previous calculations \cite{weng,feng}, since
it will only contribute a next-to-leading
additional power-law decay in the asymptotic
single electron Green's function, so one may neglect it if
only interested in the leading contribution.

The calculation for the global features of $n(k)$ is similar to
the case without string fields within MFA and the result is
\begin{eqnarray}
n(k)=1-\delta+A_{1}(k)+A_{2}(k)+A_{3}(k)+\cdot \cdot \cdot , \nonumber \\
A_{1}(k)=({2\over \pi})^{2}{\rm sin}(\delta \pi){\rm cos}(k), \nonumber \\
A_{2}(k)=({2\over \pi})^{3}[{1\over \pi}{\rm sin}^{2}(\delta \pi)
-{\delta \over 2}{\rm sin}(2\delta \pi)]{\rm cos}(2k), \nonumber \\
A_{3}(k)=({2\over \pi})^{4}[{1\over \pi}{\rm sin}(\delta \pi)
{\rm sin}(2\delta \pi)+{1\over 3}(1-{2\delta \over 3})
{\rm sin}(3\delta \pi)]{\rm cos}(3k) .
\end{eqnarray}
The curve $n(k)$ is plotted in figure 3 (solid line) for
$\delta =0.25$ in comparison with the result without string
fields (dot-dashed line). We find some substantial improvement
by including the string fields. In particular, the location
of the Fermi points was wrong in MFA without string fields,
while it is correct ($k_{F}={\pi\over 2}(1-\delta)$)
if they are included. In the
same figure we have also plotted the $n(k)$ curve obtained
in the early treatment (dashed line) \cite{feng} for CP$^{1}$
electron representation
(without accounting for {\it hard-core} nature of bosons).
Obviously, the global features of this
curve are not correct ( $n(k)$ goes up again as $k$ further increases).
Nevertheless, the asymptotic form of  the momentum
distribution near $k_{F}$,
obtained in both approaches, is correct, namely,
\begin{equation}
n(k) \sim const. -C|k-k_{F}|^{{1\over 8}}sgn(k-k_{F}) ,
\end{equation}
again in agreement with the exact numerical result \cite{ogata2}.
This singularity is washed out in the numerical calculations
and does not show up in figure 3.

To sum up we have confirmed that by introducing the string fields the
spinon-holon interactions can be included to some extent which
allows us to obtain correct exponents for the correlation
functions and momentum distribution, as proposed
earlier \cite{weng}. Moreover, the global features of the energy
spectrum and the momentum distribution, found in our fermion-spin
approach are correct in contrast with the previous
approach \cite{weng} which could not provide such an adequate
description.

\section{ The Mean Field Theory in 2D
within the Fermion-Spin Approach }

In this Section, we consider the 2D $t$-$J$ model.
Very soon after the discovery of oxide
superconductors Anderson \cite{pw1} revived his idea of the
resonating valence bond (RVB) to describe the short-range
AF fluctuations in the 2D Hubbard model.
Baskaran, Zou and Anderson \cite{pw4}
developed a mean field theory of the RVB state. Later
a number of other
more elaborated mean field solutions have been discussed both
at half filling and away from it, such as the
flux phase \cite{amk}, the spiral phase \cite{bses,dy} and the
commensurate flux
phase \cite{pw5} which breaks the time-reversal symmetry and parity.
The latter is related to the proposed fractional
statistics and anyon superconductivity \cite{rbl}.
For the early suggested flux phase \cite{amk},
the lowest energy solution is a state with ${1\over 2}$
flux quantum (or phase $\pi$) per plaquette which
does not break the time-reversal symmetry and parity. Sheng, Su and Yu
found that the $\pi$-flux phase can coexist with $s+id$ wave
RVB state for small dopings \cite{yulu6}.
Lee and Feng \cite{tkl}, and later Chen, Su, and Yu \cite{yulu2}
found that the magnetized RVB state with coexisting
AF order and a d-wave RVB order parameter, has a gain in energy.
Furthermore, Hsu \cite{hsu} has shown that the magnetized flux
state which is the coexistence of AF order with a flux state
has a similar gain in energy. Recently, Wang \cite{wang},
and Ubbens and Lee \cite{lee3},  obtained
the same result in a different framework. In
this Section, we discuss the 2D $t$-$J$
model along this line.

\subsection{Generalized Jordan-Wigner Transformation}

In the fermion-spin approach, the success of theory
depends strongly on whether one can map the
{\it hard-core boson} onto an appropriate fermion or boson
representation. In the 2D case, the Jordan-Wigner
transformation of the spin operators has been
generalized by several authors \cite{mele2}. In particular,
Wang \cite{wang}
discussed the Heisenberg model in the MFA
using this generalized Jordan-Wigner
transformation. Using our fermion-spin transformation
(7), this mean-field calculation can be easily
generalized to the $t$-$J$ model. The generalized
Jordan-Wigner transformation  may be written as
 \cite{mele2,wang}:
\begin{equation}
S^{+}_{i}=f^{\dagger}_{i}e^{i\theta_{i}},
\end{equation}
\begin{equation}
S^{-}_{i}=f_{i}e^{-i\theta_{i}},
\end{equation}
\begin{equation}
S^{Z}_{i}=f^{\dagger}_{i}f_{i}-{1\over 2},
\end{equation}
where $f_{i}$ is a spinless fermion,
$\theta_{i}=\sum_{l\neq i}f^{\dagger}_{l}f_{l}B_{il}$. In
order to preserve the {\it hard-core} properties of the spin
operators, $B_{il}$ should be such that
$e^{iB_{il}}=-e^{iB_{li}}$. One possible way of materializing
this equality is to set $B_{il}={\rm Imln}(Z_{l}-Z_{i})$, with
a complex representation of the lattice sites $Z_{l}=X_{l}+iY_{l}$.
If the effects connected with the projection operator $P$
are neglected, then
the 2D $t$-$J$ model can be mapped onto the fermion
representation as
\begin{eqnarray}
H=-t\sum_{\langle ij\rangle }h_{i}h^{\dagger}_{j}
(f^{\dagger}_{i}f_{j}
e^{i(\theta_{i}-\theta_{j})}+h.c.)
-\mu \sum_{i}h_{i}h^{\dagger}_{i} ~~~~~~~~~~~~ \nonumber \\
+J\sum_{\langle ij\rangle }h_{i}h^{\dagger}_{i}h_{j}
h^{\dagger}_{j}[{1\over 2}
(f^{\dagger}_{i}f_{j}e^{i(\theta_{i}-\theta_{j})}+h. c.)
+(f^{\dagger}_{i}f_{i}-{1\over 2})(f^{\dagger}_{j}f_{j}-{1\over 2})] .
\end{eqnarray}
It has been shown \cite{mele2,wang} that the phase factor in the
Hamiltonian (63) creates
a gauge field, with the vector potential given by
\begin{equation}
{\bf A}(r_{i})=\sum_{l\neq i}n_{l}{\hat{z}\times (r_{l}-r_{i})\over
(r_{l}-r_{i})^{2}} ,
\end{equation}
where $n_{l}=f^{\dagger}_{l}f_{l}$. On average,
there is a $\pi$-flux
attached to each spinless fermion $f_{i}$, which is nothing but
the Chern-Simons gauge field converting a {\it hard-core boson}
into a spinless fermion. As mentioned earlier,
the $\pi$-flux does not break
the time-reversal symmetry and parity.

The Hamiltonian (63) is obviously very complicated, so a more
complete discussion about it beyond the MFA
will be given elsewhere. In this Section, we only discuss some
mean-field properties of the 2D $t$-$J$ model within the
fermion-spin approach. In the MFA,
following the similar discussion of Laughlin $et~al$. \cite{rbl},
Mele \cite{mele2} and Wang \cite{wang}, the
phase factor is absorbed by redefining a bond-dependent
exchange parameter $J_{ij}$. At half filling, there are
many possible
phases and the state with the lowest energy \cite{wang}
turned out to be the magnetized flux state  with coexisting
N\' eel order parameter $\langle S^{Z}_{i}\rangle =(-1)^{i}M$ and
orbital current order parameter
$\langle S^{+}_{i}S^{-}_{i+\nu}\rangle =\chi $,
where $\nu =\pm  x, \pm  y$. The ground state energy per
bond is $E_{0}=-0.33J$, while the staggered magnetization is
$M=0.389$, which are rather close to the best numerical
estimates \cite{ceperley}
$E_{0}=-0.3346J$ and $M=0.31$, respectively.
This state is completely
equivalent to what was  first discussed by Hsu \cite{hsu},
who obtained the ground state energy $E_{0}=-0.331J$
and staggered magnetization
$M\approx 0.3$ within
a different theoretical framework. All these results show
the accuracy
of the mean-field result within the generalized Jordan-Wigner
transformation \cite{mele2}.

\subsection{Mean-Field Theory away from Half-Filling}

Away from half-filling we need to introduce an additional
holon particle-hole order parameter
$\phi =\langle h^{\dagger}_{i}h_{i+\nu}\rangle$,
and the $t$-$J$ model can be decoupled in the MFA as
\begin{eqnarray}
H=2t\chi \sum_{\langle jl\rangle}h^{\dagger}_{j}h_{l} -
\mu \sum_{j}h_{j}h^{\dagger}_{j} +
(J_{eff}+2t\phi -2J_{eff}\chi)\sum_{\langle jl\rangle}
e^{i(\theta_{j}-\theta_{l})}f^{\dagger}_{j}f_{l} \nonumber \\
-2J_{eff}M\sum_{\langle jl\rangle}(-1)^{l}f^{\dagger}_{j}
f_{j} +4NJ_{eff}\chi^{2}
-8Nt\chi \phi +4NJ_{eff}M^{2} ,
\end{eqnarray}
where $J_{eff}=J[(1-\delta)^{2}-\phi^{2}]$, and $N$ is number of
lattice sites. In the magnetized flux phase, the
$t$-$J$ model can be diagonalized by using the
Bogoliubov transformation to give
\begin{eqnarray}
H=\sum_{k (red)}(\varepsilon^{-}_{k}\alpha^{\dagger}_{k}
\alpha_{k}+\varepsilon^{+}_{k}\beta^{\dagger}_{k}\beta_{k})+
\sum_{k (red)}(E_{k}A^{\dagger}_{k}A_{k}-E_{k}B^{\dagger}_{k}
B_{k}) \nonumber \\
+4NJ_{eff}\chi^{2}-8Nt\chi \phi +4NJ_{eff}M^{2} ,
\end{eqnarray}
where (red) means the summation is carried only over
the reduced Brillouin
zone. The new operators $\alpha_{k}$, $\beta_{k}$, $A_{k}$, and
$B_{k}$ are related to $h^{A}_{k}$, $h^{B}_{k}$, $f^{A}_{k}$,
and $f^{B}_{k}$  by
\begin{equation}
h^{A}_{k}={1\over \sqrt{2}}(\alpha_{k}+\beta_{k}),~~~~~~~~
h^{B}_{k}={1\over \sqrt{2}}(\alpha_{k}-\beta_{k}),
\end{equation}
and
\begin{equation}
f^{A}_{k}={{\rm cos}k_{y}-i{\rm sin}k_{x}\over
\sqrt{{\rm cos}^{2}k_{y}+{\rm sin}^{2}k_{x}}}(u_{k}A_{k}
-v_{k}B_{k}),~~~~~~~ f^{B}_{k}=u_{k}B_{k}+v_{k}A_{k},
\end{equation}
where
\begin{equation}
u^{2}_{k}={1\over 2}(1+{8JM\over E_{k}}), ~~~~~~~~
v^{2}_{k}={1\over 2}(1-{8JM\over E_{k}}) ,
\end{equation}
and the spin excitation spectrum is
\begin{equation}
E_{k}=\sqrt{(8J_{eff}M)^{2}+4(J_{eff}+2t\phi -2J_{eff}\chi )^{2}
({\rm cos}^{2}k_{y}+{\rm sin}^{2}k_{x})} ,
\end{equation}
while the charge excitation spectrum is
\begin{equation}
\varepsilon^{\pm}_{k}=\pm 2t\chi \gamma_{k} -\mu ,~~~~~~~~~~~
\gamma_{k}=2({\rm cos}k_{x}+{\rm cos}k_{y}) .
\end{equation}
In obtaining the above results, we have considered two sublattices
$A$ and $B$ with $i \in A$ and $i+\nu \in B$.
The free energy of the 2D $t$-$J$ model can be obtained as
\begin{eqnarray}
F=-{1\over \beta}\sum_{k (red)}{\rm ln}[(1+e^{-\beta
\varepsilon^{-}_{k}})(1+e^{-\beta \varepsilon^{+}_{k}})]
-{1\over \beta}\sum_{k (red)}{\rm ln}[(1+e^{-\beta E_{k}})
(1+e^{\beta E_{k}})] \nonumber \\
+4NJ_{eff}\chi^{2}-8Nt\chi \phi +4J_{eff}M^{2} ,
\end{eqnarray}
from which we find the self-consistent equations for the
order parameters $\chi$, $M$, $\phi$ by minimizing the
free energy with respect to these parameters.

\subsection{Doping Dependence of the Staggered Magnetization}

At half-filling, the 2D $t$-$J$ model reduces to a 2D AF Heisenberg
Hamiltonian, and the result is the same as discussed by
Hsu \cite{hsu} and Wang \cite{wang}. The spinon spectrum
in the magnetized flux phase with AF long-range order is
expressed as
\begin{equation}
E_{k}=\sqrt{(8JM)^{2}+4J^{2}(1-2\chi)^{2}({\rm cos}^{2}k_{y}+
{\rm sin}^{2}k_{x})} ,
\end{equation}
where a gap appears due to the presence of AF staggered
magnetization. We
note that a gap in the spinon spectrum of the flux phase
at half filling was suggested by Laughlin earlier \cite{rbl}.
However, this gap coming
from the staggered magnetization $M$ decreases very rapidly upon
doping. In the MFA, $M$ vanishes around doping
$\delta \approx 0.1\div 0.15$ for $t/J=3\div 5$ which
is plotted in
figure 4(a). This result is in reasonably good agreement with
experiments \cite{vaknin,ykk} and
Monte-Carlo simulations \cite{tkl}. For comparison
we note that the
magnetization vanishes only at $\delta \approx 0.62$
in the MFA for the slave-fermion approach \cite{dy}.
At finite dopings, but still within the magnetized flux phase,
we find a competition between the N\' eel order
parameter $M$ and the orbital current order parameter $\chi$,
with $M$ decreasing very rapidly
(see figure 4(a)), and  $- \chi$ increasing very fast
(see figure 4(b)) upon doping. In the small doping range,
the holon particle-hole
order parameter $\phi$ increases roughly linearly upon doping,
and
is almost independent of $t/J$, which is shown in figure 4(c).
The $t$-$J$ model is characterized by a competition between
the kinetic energy ($t$) and the magnetic energy ($J$).
The magnetic energy $J$ favors an AFLRO for the spins,
whereas the
kinetic energy $t$ favors delocalization of the holes and tends
to destroy the spin AFLRO. Thus the rapid suppression
of the AFLRO upon doping means that the present mean-field
kinetic energy is better than those obtained in the slave-fermion
approach
and is closer to the optimal kinetic energy of the system. Beyond
doping $\delta \approx 0.1\div 0.15$, corresponding  to the
range of the actual high-temperature superconductors, there is no
AFLRO, but short range AF correlations persist
and the spinons are in a disordered flux phase with a
gapless spectrum,
\begin{equation}
E_{k}=2(J_{eff}+2t\phi -2J_{eff}\chi )\sqrt{{\rm cos}^{2}k_{y}
+{\rm sin}^{2}k_{x}} ,
\end{equation}
as conjectured by Anderson \cite{pw1}. This spinon
spectrum is similar to
that of the flux state discussed also at the mean-field level by
Affleck and Marston, and Kotliar \cite{amk}.

The mean-field phase boundary between the magnetized flux and
disordered flux states is, of course, at somewhat higher doping
$\delta$ than the value given by experiments and numerical
simulations.
In fact, the frustrations of spins can shift the mean-field
phase boundary towards smaller doping $\delta$ \cite{feng6}.
Thus we believe
that the result will be even closer to experiments and
numerical simulations by going beyond the MFA.

\subsection{Specific Heat}

The specific heat measurements on oxide superconductors have been
made for many compounds by different researchers \cite{hef}. The
descrepancics are mostly due to the difficulty in preparing
and characterizing samples of the oxide compounds. Although the
specific heat data for the superconducting compounds
show considerable variations
for samples measured by different groups,  some qualitative
features seem common to all the
measurements.
Hence a quantitative comparison between theory and
experiment is still early, but the qualitative tendency of the
specific heat in an adequate theoreticl description
should be consistent with experiments.

In the half filled case, there are no charge degrees of freedom
and the
spinon specific heat has been considered by Wang \cite{wang}.
Away from half-filling, we are interested in the doping range
$\delta =0.1 \sim 0.2$ where the superconductivity appears.
In this doping range, we have shown that the magnetization $M$
vanishes and the system is in a disordered flux state, where
the internal energy of the system in the MFA
is given by
\begin{equation}
U(T)=4NJ_{eff}\chi(1 -\chi^{2}) -8Nt\chi \phi ,
\end{equation}
from which the specific heat can be obtained as
\begin{equation}
C_{v}(T)=({\partial U \over \partial T})_{V} .
\end{equation}
The numerical result is shown in figure 5 at doping $\delta =0.2$
for $t/J$=3 (solid line), $t/J$=5 (dashed line),
and the shape
is similar to the experimental results \cite{hef,pgka}.
For $T>0.0005J$,
the specific heat is found to increase with temperature $T$,
which is also consistent with the experiments. Therefore our simple
mean-field calculation provides a correct qualitative description
of the specific heat for oxide compounds.

\subsection{Phase Separation}

The possible phase separation in the $t$-$J$ model
was proposed by Emery, Kivelson, and Lin \cite{ekl}.
They argued that the dilute
holes in an antiferromagnet are unstable against phase separation
into  hole-rich and  no-hole phases, and the transition from an
ordered state to doped state is of first order.
Later investigation
\cite{rice1} using high-temperature expansions shows that a line
of the phase separation extends from $J/t=3.8$ at zero filling to
$J/t=1.2$ near half-filling, but for the range of parameter
interesting to the copper oxide planes there is no evidence for
phase separation. Within the mean-field theory of our
fermion-spin approach, we find that the phase separation is
robust for the $t$-$J$ model, and the phase separation manifests
itself at the mean-field level by a negative compressibility
(slope of the chemical potential). The total energy $E_{total}$
and the chemical potential $\mu$ as a function of the
doping $\delta$
for $t/J=5$ is plotted in figure 6, which shows that the phase
separation occurs roughly at dopings $\delta\leq \delta_{c} =0.08$.
In this doping range ($\delta \leq \delta_{c}$), the compressibility
of the $t$-$J$ model is negative, and therefore the magnetized flux
state with long-range order is unstable. The range of the phase
separation will be reduced by considering the
frustrations of spins \cite{feng6}.

\section{Summary and Discussions}

In this paper, we have developed a theoretical framework,
fermion-spin
approach, to study the low-dimensional $t$-$J$ model.
In this approach,
the physical electron is decoupled as a product of a
spinless fermion (holon)
and {\it hard-core boson} (spinon). The main advantage
of this approach is that the on-site local
constraints of the $t$-$J$ model or the large $U$ limit
Hubbard model
can be treated exactly in analytical calculations.
In this framework, we have
shown that the holon behaves like a spinless fermion,
while the spinon is
neither boson nor fermion, but a {\it hard-core boson},
and the sum rule for
the physical electron is obeyed. This is not true for the
conventional slave-particle
theories, where the spinon behaves like boson (slave-boson approach),
or  fermion
(slave-fermion approach), and the sum rule for the physical
electron is not
obeyed {\it within the decoupling scheme} \cite{feng}.

We have applied
this approach to study the low-dimensional $t$-$J$ model.

In the 1D case,
we have obtained  gapless spinon
and holon spectra, a good ground state energy, and a reasonable
electron-momentum distribution within the MFA.
Thus the ground-state in the fermion-spin formalism is in some sense
closer to the Bethe-ansatz Lieb-Wu's exact wave function of the 1D
large $U$ limit Hubbard model than the slave-particle approach.
It is shown
that the spinon and consequently also the physical electron
behave like Luttinger liquids.
We have also obtained the correct asymptotic
form of the spin-spin correlation functions
as well as the electron-momentum
distribution function of the 1D $t$-$J$ model
within the fermion-spin approach
by considering the string effects, with results  in
agreement with the exact
numerical simulations of the 1D large $U$ limit
Hubbard model obtained
by Ogata and Shiba \cite{ogata2}.

The 1D problem is a good testing ground where the charge and
spin are truly separated (not in the literal sense of product
of spinon and holon, but rather as independent collective
excitations) and the Fermi liquid theory fails to describe its
physical
properties. To our knowledge, neither the standard perturbation
theory, nor the conventional slave particle approach is capable
of handling this aspect. Our results of the MFA seem to hint
that the fermion-spin approach has some potential to further explore.

The results for 2D are also very encouraging. The magnetized flux
phase with a gap proportional to the staggered magnetization,
in the spinon excitation spectrum, has the lowest energy at
half-filling in the MFA. However, this AFLRO fades away at
$10\div 15$\% hole doping for $t/J =3\div 5$, beyond which a
gapless flux state becomes stable. This result agrees with
experiments and numerical simulations \cite{vaknin,ykk,tkl}. It is
essential that this mean-field result was obtained without
any adjustable parameters. This means that the formalism itself
is powerful enough to handle the frustration (delocalization)
effect of the $t$-term in destroying the AFLRO favoured by the
$J$-term. However, we should mention that the spin excitation
spectrum at half-filling is not gapless in this approach within
the MFA as it should be (spin waves). Both Laughlin's
approach \cite{rbl} and Hus's treatment \cite{hsu}
suffer from the same weakness. Probably,
it can be cured by including vertex corrections.

As mentioned in Sec. II, this formalism doubly counts the
empty site by assigning a "spin" to it. As shown
there, this defect can be cured by introducing a projection
operator to remove the extra degrees of freedom. However, in
our mean-field treatments we have not taken this projection
operator explicitly into account. The fact, that we still
obtain very good results as summarized above, indicates that
we are not making substantial errors by allowing this extra
degree of freedom, at least in  problems considered so far.

A natural question is: What is the reason why this
simple-minded transformation is so useful. To our present
understanding, there are, at least, three reasons:
(1) The local constraint
is exactly satisfied even at the mean-field level. (2) The
{\it hard-core} nature is kept in the calculation via the
Jordan-Wigner transformation in 1D \cite{jordan} or its
generalization in 2D \cite{mele2}.
(3) The representation of the {\it hard-core} boson in terms
of spin raising and lowering operators is essential, because
whenever a hole hops it gives rise immediately to a change of
the spin background as a result of careful treatment of the
constraint given in Sec. II B.
This is why the $t$-term is so efficient
in destroying the AFLRO. Of course, there are many open
questions in this approach, {\it e.g.}, how to go beyond the MFA,
what is the gauge field in this approach and what are its
major effects and so on. These and other related issues are
under investigation now.

\vskip 3truecm

After submitting the original version of this paper,
we found Ref. 56 where a
similar approach was used to study the normal state
properties  of oxide superconductors. Apart from the
difference in issues addressed in our paper and theirs
a careful reader could easily discover the substantial
distinction in the
interpretation of transformations used in these
two papers.

\acknowledgments

The authors would like to thank P. Fazekas, R. B. Laughlin,
N. Nagaosa,
A. Parola, S. Sarker, E. Tosatti, G. M. Zhang, and X. Y. Zhang
for helpful discussions.
S. P. Feng and Z. B. Su would like to thank the ICTP
for the hospitality.

\newpage

\appendix{Matrix Representation of the Projection Operator}

The holon operators $h^{\dagger}$ and $h$ in the basis
$\left(\matrix{ 1 \cr 0 \cr} \right )_{h}$,
$\left(\matrix{ 0 \cr 1 \cr} \right )_{h}$  of holon states are
given by
\begin{equation}
h^{\dagger}=\left(\matrix{0 &1\cr 0 &0\cr} \right)_{h} ,~~~~
h=\left(\matrix{0 &0\cr 1 &0\cr} \right)_{h} ,
\end{equation}
while the spin raising and lowering operators $S^{+}$, $S^{-}$
in the spin 1/2 space
$\left(\matrix{ 1 \cr 0 \cr} \right )_{s}$,
$\left(\matrix{ 0 \cr 1 \cr} \right )_{s}$
are given by
\begin{equation}
S^{+}=\left(\matrix{0 &1\cr 0 &0\cr} \right)_{s},~~~~
S^{-}=\left(\matrix{0 &1\cr 0 &0\cr} \right)_{s} .
\end{equation}
In the product space $|hole \rangle \otimes |spin \rangle$ the
basis vectors are
\begin{eqnarray}
|1,\uparrow \rangle =\left(\matrix{ 1 \cr 0 \cr} \right )_{h}
\otimes \left(\matrix{ 1 \cr 0 \cr} \right )_{s}=
\left(\matrix{ 1 \cr 0 \cr 0 \cr 0 \cr} \right ) ,~~~
|1,\downarrow \rangle =\left(\matrix{ 1 \cr 0 \cr} \right )_{h}
\otimes \left(\matrix{ 0 \cr 1 \cr} \right )_{s}=
\left(\matrix{ 0 \cr 1 \cr 0 \cr 0 \cr} \right ) , \nonumber \\
|0,\uparrow \rangle =\left(\matrix{ 0 \cr 1 \cr} \right )_{h}
\otimes \left(\matrix{ 1 \cr 0 \cr} \right )_{s}=
\left(\matrix{ 0 \cr 0 \cr 1 \cr 0 \cr} \right ) ,~~~
|0,\downarrow \rangle =\left(\matrix{ 0 \cr 1 \cr} \right )_{h}
\otimes \left(\matrix{ 0 \cr 1 \cr} \right )_{s}=
\left(\matrix{ 0 \cr 0 \cr 0 \cr 1 \cr} \right ) ,
\end{eqnarray}
which form a complete set.

The fermion-spin transformation defined by Eq. (7) gives the
following matrix representation for the fermion operators
\begin{eqnarray}
C_{\uparrow}=h^{\dagger}S^{-}=\left(\matrix{0 &1\cr 0 &0\cr}
\right)_{h}\otimes \left(\matrix{0 &0\cr 1 &0\cr} \right)_{s}=
\left(\matrix{0&0&0&0 \cr 0&0&1&0 \cr 0&0&0&0 \cr 0&0&0&0 \cr}
\right),  \nonumber \\
C_{\downarrow}=h^{\dagger}S^{+}=\left(\matrix{0 &1\cr 0 &0\cr}
\right)_{h} \otimes \left(\matrix{0 &1\cr 0 &0\cr} \right)_{s}=
\left(\matrix{0&0&0&1 \cr 0&0&0&0 \cr 0&0&0&0 \cr 0&0&0&0 \cr}
\right), \nonumber \\
C^{\dagger}_{\uparrow}=hS^{+}=\left(\matrix{0 &0\cr 1 &0\cr}
\right)_{h}\otimes \left(\matrix{0 &1\cr 0 &0\cr} \right)_{s}=
\left(\matrix{0&0&0&0 \cr 0&0&0&0 \cr 0&1&0&0 \cr 0&0&0&0 \cr}
\right), \nonumber \\
C^{\dagger}_{\downarrow}=hS^{-}=\left(\matrix{0 &0\cr 1 &0\cr}
\right)_{h}\otimes \left(\matrix{0 &0\cr 1 &0\cr} \right)_{s}=
\left(\matrix{0&0&0&0 \cr 0&0&0&0 \cr 0&0&0&0 \cr 1&0&0&0 \cr}
\right).
\end{eqnarray}
On the other hand, there are only three physical states in the
constrained Hilbert space, namely
\begin{equation}
|0\rangle =\left(\matrix{ 1 \cr 0 \cr 0 \cr} \right ),~~~~
|\uparrow \rangle =\left(\matrix{ 0 \cr 1 \cr 0 \cr} \right ),~~~~
|\downarrow \rangle =\left(\matrix{ 0 \cr 0 \cr 1 \cr} \right ).
\end{equation}
To remove the extra degrees of freedom in the
$|hole \rangle \otimes |spin \rangle $ space, we introduce a
projection operator $P$. By requiring
$P|1,\uparrow \rangle =P|1,\downarrow \rangle =|0\rangle$,
$P|0,\uparrow \rangle =|\uparrow \rangle$, and
$P|0,\downarrow \rangle =|\downarrow \rangle $, one can easily
find its matrix representation
\begin{equation}
P=\{P_{\kappa \alpha}\}=
\left(\matrix{1&1&0&0 \cr 0&0&1&0 \cr 0&0&0&1 \cr} \right),
\end{equation}
and its hermitian conjugation
\begin{equation}
P^{\dagger}=
\left(\matrix{1&0&0 \cr 1&0&0 \cr 0&1&0 \cr 0&0&1 \cr} \right).
\end{equation}
Using this projection operator, the electron operators in the
restricted Hilbert space are given by
\begin{eqnarray}
\tilde{C}_{i\uparrow}=Ph^{\dagger}S^{-}P^{\dagger}=
\left(\matrix{0&1&0\cr 0&0&0\cr 0&0&0\cr} \right),~~~~
\tilde{C}^{\dagger}_{i\uparrow}=PhS^{+}P^{\dagger}=
\left(\matrix{0&0&0\cr 1&0&0\cr 0&0&0\cr} \right),\nonumber \\
\tilde{C}_{i\downarrow}=Ph^{\dagger}S^{+}P^{\dagger}=
\left(\matrix{0&0&1\cr 0&0&0\cr 0&0&0\cr} \right),~~~~
\tilde{C}^{\dagger}_{i\downarrow}=PhS^{-}P^{\dagger}=
\left(\matrix{0&0&0\cr 0&0&0\cr 1&0&0\cr} \right),
\end{eqnarray}
as quoted in Eq. (11). It is then straightforward
to verify the operator relations quoted in the main text
Eqs. (12) and (13). In particular, the hole number operator
\begin{equation}
n_{h}=\left(\matrix{1&0&0\cr 0&0&0\cr 0&0&0\cr} \right)=
|0\rangle \langle 0|={1\over 2}P(|1\uparrow\rangle \langle 1
\uparrow |+|1\downarrow \rangle \langle 1\downarrow |)P^{\dagger}.
\end{equation}
The physical meaning of Eq. (A9) is transparent: The empty state
should be counted only once, not twice. Since in the mean-field
treatment the constraint on average doping concentration $\delta$
is imposed directly on $h^{\dagger}h$, the sum rule for the
physical electron is satisfied.

\figure{The momentum distribution of physical electrons
in the mean-field approximation obtained by the
fermion-spin transformation
proposed in this paper (solid line) in comparison with
corresponding curves in the slave-boson (dashed line) and
the slave-fermion (dot-dashed line) approaches.  The
doping concentration $\delta =0.5$.}

\figure{ The spin background is assumed to be an antiferromagnetic
state for
the $J>0$ case of the 1D $t$-$J$ model. The $t$-$J$
chain is squeezed and the spin
configuration is rearranged due to the
spin-up electron hopping:
(a) Before hopping, when the holon is at site $i$.
(b) After hopping,
when the holon has hopped to site $i+1$ from site $i$,
while the spinon
has hopped to site $i$ from site $i+1$. The spin
polarization directions to the left of the hole are already
optimized
by the fermion-spin transformation, but there is still a
hole in the chain.
(c) Squeezing out the hole from the $t$-$J$ chain.
After this squeezing, the
spin configuration is not optimal to favor the lowest spinon energy.
(d) Rearranging the spin configuration from -$\infty$ to  site $i$
to favor the lowest spinon energy. The situation for the spin-down
electron hopping is similar.}

\figure{The momentum distribution of physical electrons
in the mean-field approximation obtained by the
fermion-spin transformation with the string fields
(solid line), and without string fields (dot-dashed line),
in comparison with the corresponding curve in the
conventional CP$^{1}$
approach (dashed line) (see Ref. 24).
The doping concentration $\delta =0.25$.}

\figure{ (a) The staggered magnetization,
(b) the orbital current order parameter $\chi$,
and (c) the order parameter $\phi$ as a function of the hole
concentration $\delta$,
for $t/J$=5 (solid line), $t/J$=3 (dashed line).
MF means magnetized flux phase with long-range order, while
DF is the disordered flux phase. }

\figure{ Specific heat data as a function of temperature
$T$ (in units of $J$)
at the hole concentration $\delta =0.2$ for $t/J=3$ (solid line),
$t/J=5$ (dashed line). }

\figure{ The total energy $E_{total}$ (dashed line) and the chemical
potential $\mu /3$ (solid line) as a function of doping $\delta$ for
$t/J=5$. The range of the phase instability is roughly
$0< \delta \leq 0.08$.}

\end{document}